%% file: main.tex
\documentclass[a4paper]{PoS}

\title{
Higher-order corrections to the splitting functions from differential equations in QCD}

\ShortTitle{Higher-order corrections to the splitting functions from differential equations in QCD}

\author{Oleksandr Gituliar\\
        Institute of Nuclear Physics, Polish Academy of Sciences \\ ul.~Radzikowskiego 152, 31-342, Krak\'ow, Poland\\
        E-mail: \email{oleksandr.gituliar@ifj.edu.pl}}

\abstract{%
We report on the status an ab initio computation of the time-like splitting functions at next-to-next-to-leading order in QCD.
Time-like splitting functions govern the collinear kinematics of inclusive hadron production in $e^+e^-$ annihilation and the evolution of the parton fragmentation distributions.
Current knowledge about them at three loops has been inferred by means of crossing symmetry from their related space-like counterparts, the deep-inelastic structure functions and parton densities.
In this approach certain parts of the off-diagonal quark-gluon splitting function are left undetermined, which calls for an independent calculation from first principles.
We outline the method for calculating master integrals from differential equations which are required to attack the problem.
\\
  \begin{flushright}
    IFJPAN-IV-2016-1 \\
  \end{flushright}
}

\FullConference{12th International Symposium on Radiative Corrections (Radcor 2015) and LoopFest XIV (Radiative Corrections for the LHC and Future Colliders)\\
     15-19 June  2015\\
     UCLA Department of Physics \& Astronomy Los Angeles, CA, USA}

\usepackage{enumitem,nccmath,pdflscape,subcaption,xspace}

\DeclareMathOperator{\Order}{\mathcal{O}}

\newcommand{\as}{\alpha_s}
\newcommand{\D}{\mathrm{d}}
\newcommand{\eps}{\epsilon}
\newcommand{\FIRE}{\texttt{FIRE}\xspace}
\newcommand{\LiteRed}{\texttt{LiteRed}\xspace}
\newcommand{\Reduze}{\texttt{Reduze}\xspace}
\renewcommand{\sp}[2]{\,#1\!\cdot\!#2\,}

\begin{document}

\section{Introduction}

Splitting functions play a key role in QCD. 
As defined in the collinear factorization formalism, they let to precisely calculate fragmentation and parton distribution functions by solving DGLAP evolution equations at higher orders in pQCD.
It is also very hard to underestimate their role in calculating real contributions to the hard processes in various subtraction schemes when the need to integrate local counter-terms arise \cite{GGG05,STD06,Cza10}.
Space-like splitting functions, responsible for the evolution of parton distribution functions, are known exactly up to next-to-next-to-leading order (NNLO), i.e., $\Order(\as^3)$ in QCD \mbox{\cite{MVV04a,MVV04b}}.
On the other hand, their time-like analogues, responsible for the evolution of fragmentation functions, are not known exactly at NNLO \cite{MMV06,MV08,AMV11}.
In particular, certain parts of the off-diagonal quark-gluon time-like splitting function are left undetermined \cite{AMV11}, which calls for an independent calculation from first principles.

We have already described in \cite{GM15} how to calculate these unknown parts from the annihilation process $e^+ e^- \to \gamma^* \to \text{partons}$.
In this paper we continue our research and discuss a new method~\cite{Git15} for calculating master integrals for splitting functions from differential equations.
In the recent years, the method of differential equations became very popular due to the fact that a good choice of the basis for master integrals leads to significant simplifications of the differential equations \cite{Henn13}.
In general, it is not easy to find an appropriate basis, however a series of methods exist to solve that problem.
For example, approaches based on the Moser algorithm \cite{Mos60} discussed in~\cite{BP07,BP09,Henn14} allow to reduce a Poincar\'e rank of the system in all singular point, except maybe one (usually chosen to be infinity).
These algorithm are implemented in \texttt{Maple} as \texttt{DEtools} (\texttt{moser\_reduce} and \texttt{super\_reduce} functions) and \texttt{ISOLDE} \cite{BPS13} packages.
A significant improvement of the Moser algorithm was presented in \cite{Lee14}.
It allows to adjust transformations of the system in such a way that they do not spoil behavior in any other singular point.
Unfortunately, a computer implementation of this method, very desirable to automatize a calculation process, is unavailable yet.
For more examples of the related methods see \cite{GMTW14,MSZ14,Tan15,MO15}.

In this report we overview an alternative method \cite{Git15} for calculating phase-space master integrals with $x$-space projection from differential equations.
This method is relatively simple and can be easily implemented as computer code in order to automatize a calculation process.
At the same time it gives a complete solution for masters to any power in the dimensional regulator $\eps$ and can be successfully applied for calculating splitting functions, but is not limited to that case only.
To illustrate the method we consider master integrals which appear in the NLO contribution to the off-diagonal time-like splitting functions and discuss possible extensions to NNLO accuracy.

\section{Master integrals from differential equations}
\label{sec:deq}

Let us consider a homogeneous system of differential equations, which takes the following form
\begin{equation} \label{eq:f}
  \frac{\partial f_i}{\partial x} = \sum_{j=1}^{n} a_{ij}(x,\eps) \, f_j(x,\eps),
\end{equation}
where the coefficients $a_{ij}(x,\eps)$ of the $n\times n$ matrix $A(x,\eps)$ are known, $f_i(x,\eps)$ are unknown functions, and $\eps$ is an infinitesimally small parameter (playing the role of a dimensional regulator in $m=4-2\eps$ dimensions).

Assuming that the coefficients $a_{ij}(x,\eps)$ are rational functions of $\eps$, without loss of generality we can write
\begin{equation} \label{eq:a_eps}
  A(x,\eps) = \sum_{k=r_\eps}^\infty \eps^k \, A^{(k)}(x),
\end{equation}
where $r_\eps$ is an integer number and denotes, as we call it, an {\em $\eps$-rank} of the matrix $A(x,\eps)$.

At the same time, we restrict the matrix $A(x,\eps)$ to be of the form
\begin{equation} \label{eq:a_x}
  A(x,\eps) = \sum_{i} \frac{A_i(x,\eps)}{(x-x_i)^{1-p_i}},
\end{equation}
where $p_i$ is said to be the {\em Poincare rank} of $A(x,\eps)$ at a singular point $x_i$, and components of $A_i(x,\eps)$ are regular polynomials at $x=x_i$.
Such a form is imposed exclusively for a practical reason since calculations of the splitting functions are bound to the case of $x_i \in \{-1,0,1\}$, which is exactly an alphabet for Harmonic Polylogarithms (HPLs)~\cite{RV99}.
In the case of a more complex structure of denominators in the expansion \eqref{eq:a_x} the same arguments could be extended to the more general case of Generalized Harmonic Polylogarithms (GHPLs) introduced in~\cite{AB04}, which maintain the structure and properties of HPLs~\cite{BDV10,ABS13}.

Keeping all the above considerations in mind we proceed with solving~\eqref{eq:f} as a Lorentz series in $\eps$.
Taking into account a recursive definition of (G)HPLs we show that such a series can be found to any order in~$\eps$ at a low computational price.

\subsection{Solutions for $\eps$-rank $>0$} \label{sec:eps0}

Let us consider solution of the system \eqref{eq:f} in the form
\begin{equation} \label{eq:fi}
  f_i(x,\eps) = \sum_{k=1}^\infty \eps^k f_i^{(k)}(x).
\end{equation}
Taking into account eq.~\eqref{eq:a_eps} it is easy to show that functions $f_i^{(k)}(x)$ in \eqref{eq:fi} can be calculated by the recursive formula
\begin{equation} \label{eq:f_sol}
  f_i^{(k)}(x) = c_i^{(k)} + \sum_{m=1}^k \int \D x \, a_{ij}^{(m)}(x) f_j^{(k-m)}(x),
\end{equation}
where $c_i^{(k)}$ are integration constants determined from boundary conditions as described in Section~\ref{sec:boundary}.

\subsection{Solutions for $\eps$-rank $=0$}

Solutions for the system with $\eps\text{-rank}=0$ can not be found in the general case.
For some special cases, like {\em weakly coupled} systems, it is possible to write such solutions.
In this work we consider only this type of systems.
Weakly coupled are systems for which $a_{ij}^{(0)}(x)$ is a triangular matrix, i.e.,
\begin{equation}
  a_{ij}^{(0)}(x) = 0, \quad\text{ for }\quad i<j.
\end{equation}
In this case we can find a basis so that a new system has $\eps$-rank $>0$ and can be solved using the method of Section~\ref{sec:eps0}.
In the remaining part of this section we provide a procedure how to increase an $\eps$-rank of the system, which consists of two steps: find new bases such that
\begin{enumerate}[label=\roman*)]
  \item diagonal elements of $a_{ij}^{(0)}(x)$ are zero, i.e., $a_{ij}^{(0)}(x)=0$ for $i=j$; and
  \item off-diagonal elements of $a_{ij}^{(0)}(x)$ are zero, i.e., $a_{ij}^{(0)}(x)=0$ for $i>j$.
\end{enumerate}
These conditions ensure that for the new basis $a_{ij}^{(0)}(x) = 0$, which guarantees that $\eps$-rank $> 0$.

\subsubsection{Zero-diagonal form}
\label{sec:321}

It is easy to verify that a system of differential equations for a new basis defined as
\begin{equation} \label{eq:g-f}
  g_i(x,\eps) = b_{i}(x,\eps) f_i(x,\eps),
\end{equation}
where
\begin{equation}
  b_{i}(x,\eps) = \exp\left(-\int \D x \; a_{ii}(x,\eps)\right),
\end{equation}
contains zeroes as diagonal elements and has a new form
\begin{equation} \label{eq:g}
  \frac{\partial g_i}{\partial x} = \sum_{j=1}^{n} \tilde{a}_{ij}(x,\eps) \, g_j(x,\eps), \quad \text{where} \quad \tilde{a}_{ij}(x,\eps) = \frac{a_{ij}(x,\eps)}{b_{j}(x,\eps)}.
\end{equation}

\subsubsection{Zero-triangular form}
\label{sec:322}

Next, following the same strategy, we find a new basis $h_i(x,\eps)$ which leads to the zero-triangular form of the differential equations:
\begin{equation} \label{eq:h_ij}
  h_i(x,\eps) = g_i(x,\eps) + \sum_{j=1}^{i-1} b_{ij}(x,\eps) g_j(x,\eps),
\end{equation}
where
\begin{equation} \label{eq:h_b_ij}
  b_{ij}(x,\eps) = - \int \D x \bigg( \tilde{a}_{ij}^{(0)}(x) + \sum_{k=j+1}^{i-1} b_{ik}(x,\eps) \tilde{a}_{kj}^{(0)}(x) \bigg).
\end{equation}
A complete form of the new system is rather complex and it is of no practical use to write it down here. However it can be easily obtained from \eqref{eq:h_ij} after coefficients of \eqref{eq:h_b_ij} are explicitly calculated.

Let us show that such a choice indeed leads to the desired zero-triangular system of equations.
Taking the derivative of \eqref{eq:h_ij}, keeping in mind \eqref{eq:g} and \eqref{eq:h_b_ij}, and neglecting higher-order term in $\eps$ we obtain
\begin{equation} \label{eq:part_h}
  \frac{\partial h_i}{\partial x}
  =
 \sum_{j=1}^{i-1}
 \bigg(
   \tilde{a}_{ij}^{(0)} g_j - \bigg( \tilde{a}_{ij}^{(0)} g_j + \sum_{k=j+1}^{i-1} b_{ik} \tilde{a}_{kj}^{(0)} g_j \bigg) +\sum_{k=1}^{j-1} b_{ij}  \tilde{a}_{jk}^{(0)} g_k
 \bigg)
 .
\end{equation}
It is easy to check, by carefully switching summation variables in one of the nested sums, that right-hand side of \eqref{eq:part_h} becomes zero.

At first sight, it may look that nested integrals in \eqref{eq:h_b_ij} are way too complicated for practical calculations.
In fact, they are very easy to compute taking into account the recursive nature of (G)HPLs, as was discussed earlier in this section.
For our examples, discussed in the next section, we have used the \texttt{HPL} package \cite{Maitre05}.

\subsection{Solutions for $\eps$-rank $<0$}
\label{sec:33}

As a rule, when one chooses a basis of master integrals as provided directly by the IBP rules generator, like \FIRE \cite{Smi08}, \Reduze \cite{MS12}, or \LiteRed \cite{Lee12,Lee13}, the system \eqref{eq:f} has a negative $\eps$-rank.
In this situation we can not proceed with the procedure described before in this section.
To overcome this issue it is usually enough to adjust $\eps^{n}$ factors in the masters.

To get a hint on how to choose $n$ we analyze Mellin moments for the corresponding masters that leads to several possibilities:
\begin{enumerate}
  \item In the presence of factors $x^{-1+a\eps}(1-x)^{-1+b\eps}$ we choose $n = r_\eps^{(1)} - 1$, where $r_\eps^{(i)}$ is an $\eps$-rank of the $i^\text{th}$ Mellin moment.
  The reason is that the logarithmic singularity in $x$ is canceled by the Mellin moment while the second one in $1-x$ introduces an additional $\eps$ pole.
  \item In the presence of factors $x^{-1+a\eps}$ we choose $n = r_\eps^{(0)} - 1$.
  \item Otherwise we choose $n = r_\eps^{(0)}$.
\end{enumerate}

\subsection{Boundary conditions} \label{sec:boundary}

The final step of the method is to fix boundary conditions, i.e., to find integration constants $c_i^{(k)}$ which appear in \eqref{eq:f_sol}.
On the one hand, in the case of phase-space integrals we can do that by calculating Mellin moments of the solution \eqref{eq:f_sol}.
On the other hand, the same moments can be taken from the literature or directly calculated by performing integration over the entire $n$-particle phase-space, i.e.,
By analogy with phase-space integrals with $x$-space projection, we generate IBP rules for the inclusive integrals as well.
This way we reduce the set of inclusive master integrals which should be calculated explicitly.

Another simplification is related to the Mellin moments, which can be extracted from the difference equations.
These equations in turn can be derived from the differential equations \eqref{eq:f}, hence only one Mellin moment needs to be computed for each inclusive master.

\section{NLO Master Integrals}

\begin{figure}[h]
  \centering
  \begin{subfigure}[b]{0.250\textwidth}
    \includegraphics[width=\textwidth]{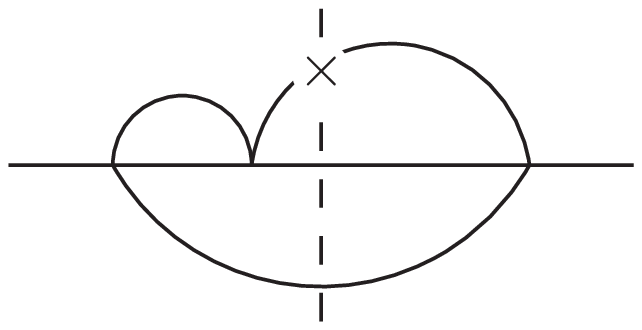}
    \caption*{$V_1 = \{1,2\}$}
  \end{subfigure}%
  ~
  \begin{subfigure}[b]{0.250\textwidth}
    \includegraphics[width=\textwidth]{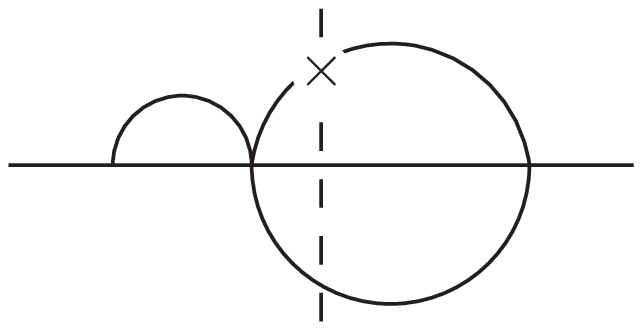}
    \caption*{$V_2 = \{1,3\}$}
  \end{subfigure}
  ~
  \begin{subfigure}[b]{0.250\textwidth}
    \includegraphics[width=\textwidth]{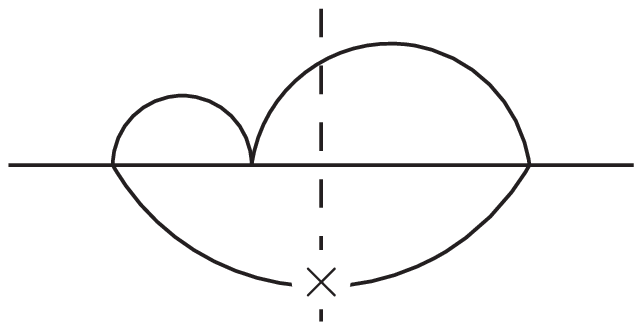}
    \caption*{$V_3 = \{1,4\}$}
  \end{subfigure}
  \vspace{4mm}

  \begin{subfigure}[b]{0.250\textwidth}
    \includegraphics[width=\textwidth]{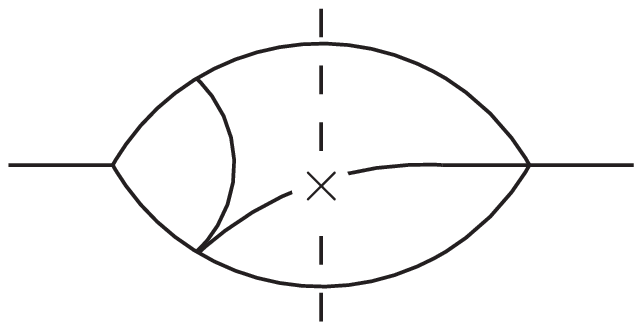}
    \caption*{$V_4 = \{1,2,3\}$}
  \end{subfigure}
  ~
  \begin{subfigure}[b]{0.250\textwidth}
    \includegraphics[width=\textwidth]{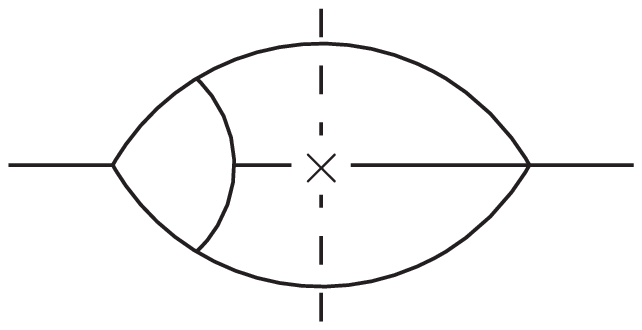}
    \caption*{$V_5 = \{1,2,3,5\}$}
  \end{subfigure}
  ~
  \begin{subfigure}[b]{0.250\textwidth}
    \includegraphics[width=\textwidth]{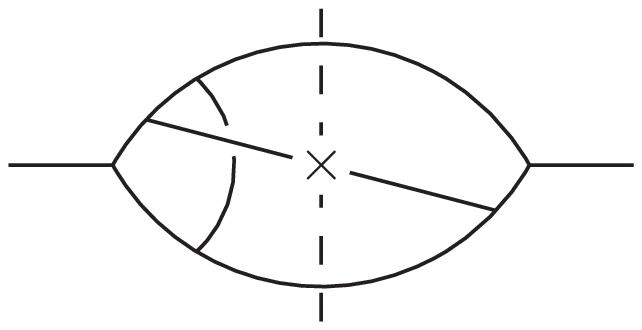}
    \caption*{$V_6 = \{1,2,3,6,7\}$}
  \end{subfigure}
  \vspace{4mm}
  \caption{Master integrals for the real-virtual NLO contribution to the time-like splitting function.}
  \label{fig:1}
\end{figure}

To illustrate our method we consider NLO contributions to the off-diagonal time-like splitting function.
We define real-virtual master integrals depicted in figure~\ref{fig:1} as
\begin{equation}
  V_i(x,\eps)
  =
  \{a_1,\ldots,a_n\}
  =
  \int \D \mathrm{PS}(2) \, \D l
    \frac{1}{D_{a_1} \ldots D_{a_n}}
,
\end{equation}
where phase-space integration with $x$-projection is defined as
\begin{equation} \label{eq:psn}
    \int \D \mathrm{PS}(n)
    =
    \int \prod_{i=0}^{n} \D^m\! k_i \; \delta^+(k_i^2) \; \delta\Big(x-\frac{2\sp{q}{k_0}}{q^2}\Big) \; \delta\Big(q-\sum_{j=0}^{n} k_j\Big)
,
\end{equation}
$l$ is a loop momentum, and denominators $D_j$ are defined in as

\begin{equation}
\begin{aligned}
  D_1 & = l^2
  &
  D_2 & = (l+k_1-q)^2
  &
  D_3 & = (l-q)^2 
  &
  D_4 & = (l+k_1+k_2)^2 
  \\
  D_5 & = (l-k_2)^2
  &
  D_6 & = (l+k_1+k_2-q)^2 
  &
  D_7 & = (k_2-q)^2.
\end{aligned}
\label{eq:Drv}
\end{equation}

\begin{figure}[h]
  \centering
  \begin{subfigure}[b]{0.200\textwidth}
    \includegraphics[width=\textwidth]{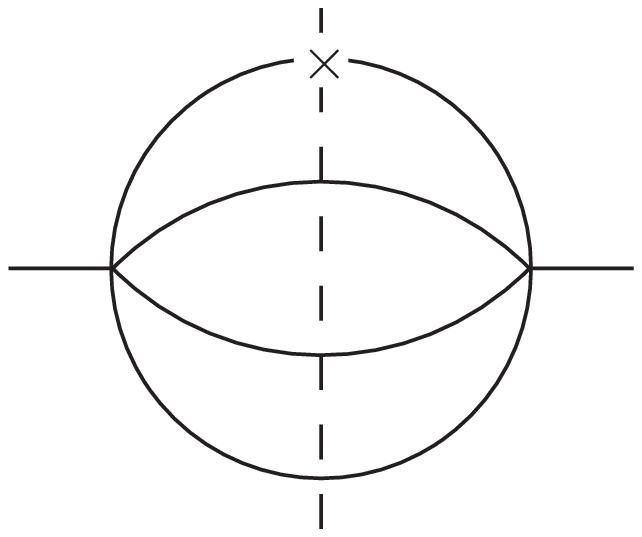}
    \caption*{$R_1 = \{\}$}
  \end{subfigure}%
  ~
  \begin{subfigure}[b]{0.200\textwidth}
    \includegraphics[width=\textwidth]{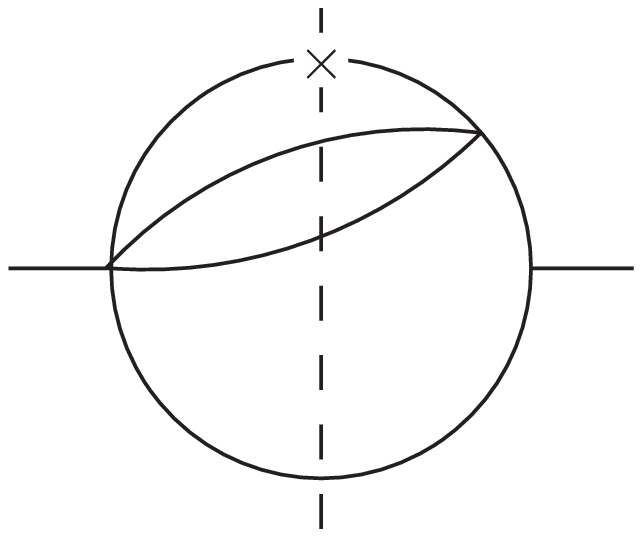}
    \caption*{$R_2 = \{2\}$}
  \end{subfigure}
  ~
  \begin{subfigure}[b]{0.200\textwidth}
    \includegraphics[width=\textwidth]{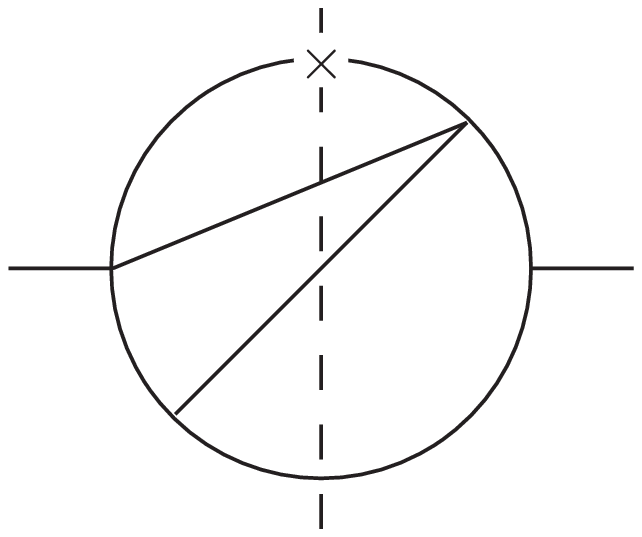}
    \caption*{$R_3 = \{3,6\}$}
  \end{subfigure}
  ~
  \begin{subfigure}[b]{0.200\textwidth}
    \includegraphics[width=\textwidth]{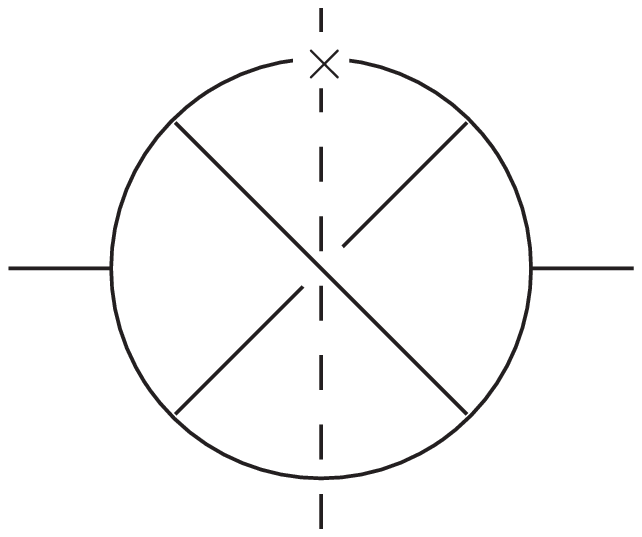}
    \caption*{$R_4 = \{4,5,6,7\}$}
  \end{subfigure}
  \vspace{4mm}

  \begin{subfigure}[b]{0.200\textwidth}
    \includegraphics[width=\textwidth]{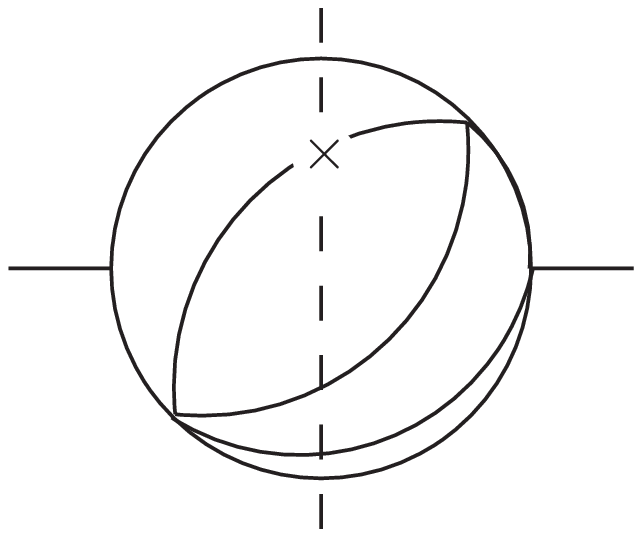}
    \caption*{$R_5 = \{1,2,3\}$}
  \end{subfigure}
  ~
  \begin{subfigure}[b]{0.200\textwidth}
    \includegraphics[width=\textwidth]{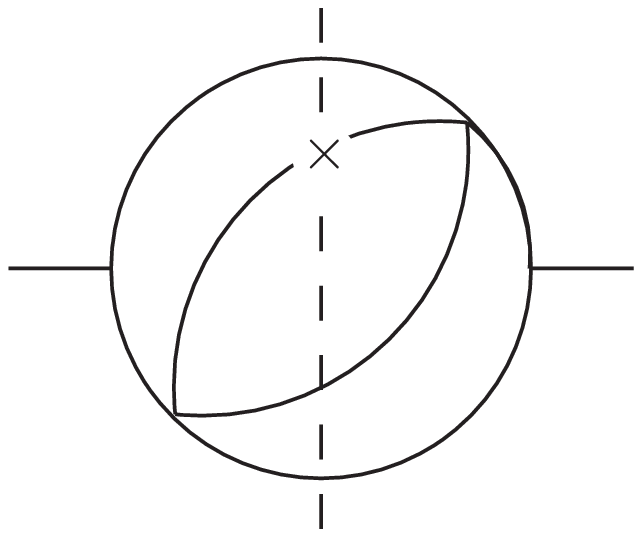}
    \caption*{$R_6 = \{2,3\}$}
  \end{subfigure}
  ~
  \begin{subfigure}[b]{0.200\textwidth}
    \includegraphics[width=\textwidth]{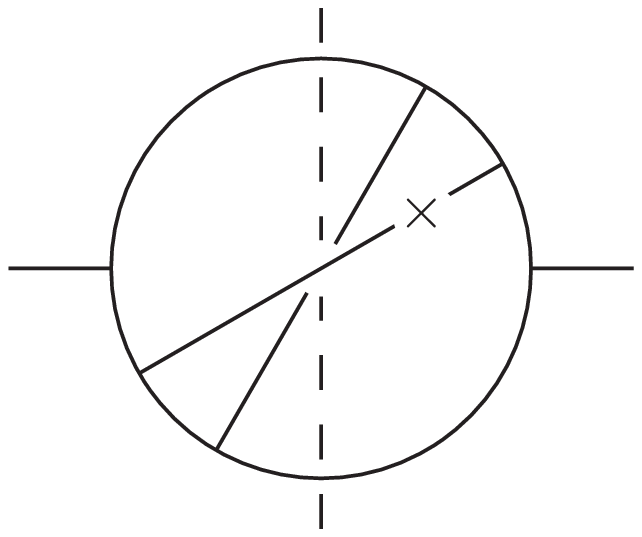}
    \caption*{$R_7 = \{2,3,6,7\}$}
  \end{subfigure}
  ~
  \begin{subfigure}[b]{0.200\textwidth}
    \includegraphics[width=\textwidth]{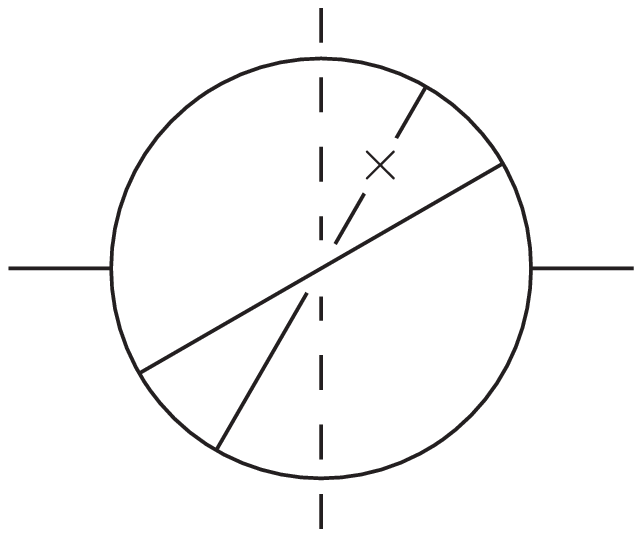}
    \caption*{$R_8 = \{2,3,4,5\}$}
  \end{subfigure}
  \vspace{4mm}
  \caption{Master integrals for the real-real NLO contributions to the time-like splitting function.}
  \label{fig:2}
\end{figure}

By analogy with the real-virtual case we proceed with the real-real contribution.
We define master integrals depicted in figure~\ref{fig:2} as
\begin{equation}
  R_i(x,\eps)
  =
  \{a_1,\ldots,a_n\}
  =
  \int \D \mathrm{PS}(3)
    \frac{1}{D_{a_1} \ldots D_{a_n}}
,
\end{equation}
where denominators $D_j$ are defined in \eqref{eq:Drr}.
\begin{equation}
\label{eq:Drr}
\begin{aligned}
  D_1 & = k_1^2 
  &
  D_2 & = (q-k_1)^2
  &
  D_3 & = (q-k_2)^2
  &
  D_4 & = (q-k_1-k_3)^2 
  \\
  D_5 & = (q-k_2-k_3)^2
  &
  D_6 & = (k_2+k_3)^2 
  &
  D_7 & = (k_1+k_3)^2
\end{aligned}
\end{equation}

Defined in such a way masters can be easily found with the help of the method described in Section~\ref{sec:deq}.
For more details of the computation we refer the reader to our work \cite{Git15} where we presented all the technical details.

\section{Summary and Prospects}

In this report we discuss a method for calculating phase-space integrals for the off-diagonal time-like splitting functions based on the differential equations method.
This work is a part of the ongoing ab initio computation of the time-like splitting functions at next-to-next-to-leading order in QCD.
The key idea of our method is to find an appropriate basis for master integrals which leads to significant simplification of differential equations.
As a main result, we describe how to efficiently construct such a basis and find solutions of the resulting differential equations.
We also discuss how to fix boundary conditions from the Mellin moments.
To demonstrate how our method works in practice, we calculate master integrals for the decay processes $1 \to 4$ and $1 \to 3$ with an $x$-space projection, needed to extract NLO contributions to the off-diagonal time-like splitting function from $e^+e^-$ annihilation process.

\acknowledgments

I gratefully acknowledge the hospitality of the Theory Group of the University of Hamburg where major part of this research was done.
In particular, I am thankful for numerous discussions with Sven-Olaf Moch, his support and guidance.

This work has been supported by the Research Executive Agency (REA) with the European Union grant PITN-GA-2010-264564 (LHCPhenoNet) and by Narodowe Centrum Nauki with the Sonata Bis grant DEC-2013/10/E/ST2/00656.

The Feynman diagrams were drawn with the help of \texttt{JaxoDraw}~\cite{BCKT08} and \texttt{Axodraw}~\cite{Ver94}.

\bibliographystyle{JHEP}

\input{main.bbl}

\end{document}

%% file: main.bbl
\providecommand{\href}[2]{#2}\begingroup\raggedright\endgroup